\documentclass[12pt]{article}
\usepackage{axodraw}

\def\thefootnote{\fnsymbol{footnote}}

\begin{document}

\begin{titlepage}
\noindent 
\begin{minipage}[t]{6in} 
\begin{flushright} 
 KEK-TH-724       \\
 UT-916           \\
 hep-ph/0011141    \\ 
\vspace*{1cm} 
\end{flushright} 
\end{minipage}

\begin{center}
{\Large
{\bf Supersymmetric seesaw model for the}                    \\
{\bf (1+3)-scheme of neutrino masses}}                       \\[4ex]
{\bf 
 F. Borzumati$^{\,a}$, K. Hamaguchi$^{\,b}$, and 
 T. Yanagida$^{\,b,c}$}                                 \\[2ex]
${}^{a}$ 
{\it Theory Group, KEK, Tsukuba, Ibaraki 305-0801, Japan}    \\
${}^{b}$ 
{\it Department of Physics, University of Tokyo,
     Tokyo 113-0033, Japan}                                  \\ 
${}^{c}$ 
{\it Research Center for the Early Universe,
     University of Tokyo, Tokyo 113-0033, Japan}             \\[12ex]
\end{center}
 
{\begin{center} ABSTRACT \end{center}}
\vspace*{1mm}

\parbox{14.5cm}
{
\noindent
A four-neutrino spectrum with a sterile neutrino without significant
involvement in the atmospheric and solar neutrino oscillation
experiments has been recently advocated as the correct picture to
explain all existing experimental data. We propose a supersymmetric
model in which this picture can be naturally implemented. In this
model, the mass for the mainly active neutrino eigenstates is induced
by the seesaw mechanism with a large intermediate scale $M_R$, whereas
the mass for the mainly sterile neutrino state is closely related to
supersymmetry breaking.
}
\vfill

\end{titlepage}

\newpage  
\setlength{\parskip}{1.01ex}

\renewcommand{\thefootnote}{\arabic{footnote}}
\setcounter{footnote}{0}

It has been recently proposed to link the mechanism for generating
small neutrino masses to the mechanism of supersymmetry
breaking~\cite{Borzumati:2000mc,Arkani-Hamed:2000bq,Arkani-Hamed:2000kj,
Babu:2000hb}. The hierarchy between the gravitino mass $m_{3/2}$ 
($\simeq 1\,$TeV) and
the Planck mass $M_P$ ($\simeq 2\times 10^{18}\,$GeV) 
induced by the breaking of supersymmetry at 
$m_X < M_P$, i.e. $m_{3/2}\sim m_X^2/M_P \ll M_P$, 
make natural the presence of small
dimensionless and dimensionful parameters such as $m_{3/2}/M_P$,
$m_{3/2}^2/M_P$, as well as $(m_{3/2}\, v) /M_P$ ($v$ is here a
typical vacuum expectation value ({\it vev}) of Higgs doublets), or,
generically $\tilde{m}/M_P$ where $\tilde{m}$ is a soft supersymmetry
breaking mass. Depending on the value of $m_{3/2}$ and of the soft
massive couplings, these parameters may be very important in the
neutrino sector, which requires the smallest masses in the spectrum of
known particles.

The by now ``standard'' seesaw 
mechanism~\cite{Yanagida:1980xy,Gell-Mann:1980vs} accounts for the
lightness of the mainly active neutrino states $\nu_1$, $\nu_2$, and
$\nu_3$ in the same way, i.e. making use of a natural suppression
factor $v^2/M_R$, where $M_R$ is an intermediate large scale. This is
usually dynamically motivated as the scale of an additional gauge
interaction or as the scale of Grand Unified Theories (GUTs). It is
generally believed that this mechanism alone is incompatible with an
additional light neutrino state with main component sterile,
i.e. insensitive to the Standard Model (SM) gauge interactions.

The sterile neutrino needed to simultaneously explain LSND and the
solar and atmospheric neutrino experiments~\footnote{It has been
 recently pointed out that all existing experimental data on neutrino
 oscillations can be explained without introducing a sterile
 neutrino, if $CPT$ is broken~\cite{Murayama:2000hm}.} is in 
the range $m_\nu \sim 10^{-4}\,$eV--$1\,$eV, if cosmological
constraints on neutrino
masses~\cite{Bond:1980ha,Gawiser:1998zh,Peebles:1993xt} are also kept
into account. Two four-neutrino pictures exist to describe all the
oscillation data. One is the well-known ``$2\!+\!2$''picture with two
pairs of neutrinos separated by a gap of squared mass $\sim 1\,$eV$^2$
($\equiv \Delta m^2_{\rm LSND}$) and with the two components of each
pair separated by the two squared mass splittings needed for the solar
and atmospheric neutrino oscillations, $\Delta m^2_{\rm atm}$ and
$\Delta m^2_{\rm sol}$~\cite{Barger:2000hs,Kayser:2000ka}. In
particular, it is 
$\Delta m^2_{\rm atm}=10^{-3}\!-\!10^{-2}\,$eV$^2$~\cite{Barger:2000hs}
for the atmospheric neutrino oscillation and 
$\Delta m^2_{\rm sol}\sim 10^{-5}\,$eV$^2$ or 
$\Delta m^2_{\rm sol}=10^{-9}\!-\! 10^{-7}\,$eV$^2$, if the MSW
oscillation or the quasi-vacuum oscillation are the correct solutions
to the solar neutrino problem~\cite{Gonzalez-Garcia:2000sk}. This
picture requires a large involvement of the sterile neutrino component
in the solar neutrino oscillation experiment, which is still 
marginally allowed 
by the Super-Kamiokande data~\cite{SK-SOL-NU2000}, although at
different levels of confidence in different analyses
(cfr.~\cite{Gonzalez-Garcia:2000sk} and~\cite{SK-SOL-NU2000}). The
second picture, recently motivated by the smaller oscillation
probability now claimed by the LSND experiment~\cite{NEWLSND}, is the
``$1\!+\!3$'' picture with one neutrino heavier than the other three
by the amount 
$\vert \Delta m^2_{\rm LSND}
\vert^{1/2}$~\cite{Kayser:2000ka,Barger:2000ch,Peres:2000ic}. The
heavier neutrino is mainly composed by the sterile neutrino; the other
three lighter states, mainly active, have masses compatible with any
of the two pair of values $(\Delta m^2_{\rm atm},\Delta m^2_{\rm
sol})$ given above. This type of spectrum favours small mixing angles
between the lighter neutrinos and the heavier one and therefore a
marginal involvement of the sterile neutrino in the solar and
atmospheric neutrino oscillations.

Light sterile neutrinos can be implemented in the scenarios proposed
in Refs.~\cite{Borzumati:2000mc} and~\cite{Babu:2000hb}.  Both
proposals rely on a specific class of models of supersymmetry
breaking~\cite{Izawa:1996pk,Intriligator:1996pu}, based on a
supersymmetric $SU(2)$ gauge theory with strong coupling. In these
models, a SM singlet $Z$, with a supersymmetry-breaking {\it vev}
${\cal F}_Z$, acquires also a {\it vev} ${\cal A}_Z$, supersymmetry
conserving, but induced by the breaking of
supersymmetry~\cite{Izawa:1995jg,Chacko:1998si}.  This singlet $Z$
couples to sterile neutrinos via non-renormalizable Yukawa-type
interactions, which give rise to small tree-level neutrino masses.

The two proposals, however, differ in some fundamental points. In the
first one~\cite{Borzumati:2000mc}, the only sterile states are the
neutrino superfields $\bar{N}$ participating in the generation of
masses for $\nu_1$, $\nu_2$, and $\nu_3$. Tree-level Yukawa couplings
for the $\bar{N}$ are forbidden by some discrete symmetry
under which the relevant superfields are charged.  Neutrino masses are
generated at the tree level by non-renormalizable operators and
radiatively. Both mechanisms, which become possible after the breaking
of supersymmetry, induce suppression factors of seesaw type, but
in general tend to 
replace the conventional seesaw mechanism for generating the mass of
$\nu_1$, $\nu_2$, and $\nu_3$. The resulting spectra may encompass the
typical seesaw spectrum with three heavy Majorana mass eigenstates 
$n_1$, $n_2$, and $n_3$, if the $\bar{N}$'s are allowed to be heavy,
or may be composed of six light eigenstates ($\nu_i$,$n_i$) for
$i=1,2,3$, if the existing discrete symmetries suppress the tree-level
mass for the $\bar{N}$ neutrinos. The exact value of masses and mixing
angles depends on the specific values of dimensionless couplings and
soft supersymmetry-breaking parameters.

In the second mechanism~\cite{Babu:2000hb}, a fourth sterile neutrino
superfield $S$ is added to the three heavy right-handed neutrinos,
which together with the three active ones, induce Majorana masses for
$\nu_1$, $\nu_2$, and $\nu_3$ through the usual seesaw mechanism. The
three Majorana states $n_1$, $n_2$, and $n_3$, are at the intermediate
scale $M_R$.  The fourth eigenstate $\nu_s$ has a mass mainly given by
the Dirac mass obtained at the tree-level from the non-renormalizable
operator $Z S LH$.  Care has to be taken for the coupling of this
interaction not to be too large, to avoid instabilities and dangerous
vacua lower than the electroweak one. Moreover, $R$-charges have to be
assigned to the relevant fields in order to avoid large radiative
contributions to the Majorana mass of active neutrinos. Further
discussion on these aspects can be found in Ref.~\cite{BHNY}.  The
light eigenstate $\nu_s$, of Dirac type, has a mass of order
$10^{-4}\,$eV, i.e. the mass required for the solar neutrino
quasi-vacuum oscillation. For $m_{\nu_2}$ and $m_{\nu_3}$ of order
$0.1\,$eV, and $m_{\nu_1}$ also of order $10^{-4}\,$eV, which can be
easily disposed for by the seesaw mechanism, this spectrum is
consistent with the ``$2\!+\!2$''picture of neutrino masses.

Motivated by the new LSND claim and the analysis of
Refs.~\cite{Barger:2000ch,Peres:2000ic}, we try now to build a model
in which the ``$1\!+\!3$'' picture can be accommodated. As in the
proposal of Ref.~\cite{Babu:2000hb}, we assume the usual seesaw
mechanism to be the one inducing small values of $m_{\nu_1}$,
$m_{\nu_2}$, and $m_{\nu_3}$ and only one sterile neutrino $S$ is
added to the three heavy $\bar{N}$'s. Once again, the mechanism for
generating $m_{\nu_s}$ is linked to supersymmetry breaking, but
differently than in the proposals of Refs.~\cite{Borzumati:2000mc}
and~\cite{Babu:2000hb}, it is independent of the specific realization
of this breaking.

Seven neutrino chiral superfields are present in this model: the three
neutral components of the leptonic doublets $L_\alpha$, with
$\alpha=e,\mu,\tau$, the three superfields $\bar{N}_\alpha$, and $S$. A
continuous $U(1)_R$ symmetry, which is known to be important for the
solution of the $\mu$ problem~\cite{Giudice:1988yz}, is assumed. Under
this symmetry the relevant SM fields and the sterile neutrino $S$ have
the following $R$-charges:
\begin{equation}
  R(L_\alpha)      =  1, \quad  
  R(\bar{N}_\alpha)=  1, \quad 
  R(H)             =  0, \quad 
  R(\bar{H})       =  0, \quad 
  R(S)             = -1\,.   
\label{rcharges}
\end{equation}
Tree-level Yukawa interactions for the sterile neutrino $S$ are
consequently forbidden.  The superpotential allowed by this symmetry
can then be decomposed as:
\begin{equation}
 W = W_0 \, + \, W_1 \, + \, W_2 \,.
\end{equation}
$W_0$ collects the usual SM Yukawa operators. $W_1$ contains mass and
interaction terms for the right-handed neutrinos $\bar{N}$:
\begin{equation}
 W_1 =  y_{\alpha \beta}               \bar{N}_\alpha L_\beta  H  \, + \,
       \frac{1}{2} M_{R\,\alpha \beta} \bar{N}_\alpha \bar{N}_\beta \,,
\label{spotone}
\end{equation}
where $M_{R\,\alpha \beta}$ are the typical large seesaw masses.
Finally, $W_2$ contains all mass and interaction terms for $S$,
together with an operator giving rise to the bilinear $\mu H \bar{H}$
term:
\begin{equation}
 W_2 =  
  h_H          H \bar{H}        \frac{\langle W\rangle}{M_P^2}   
\, + \,
  f_\alpha     \bar{N}_\alpha S \frac{\langle W\rangle}{M_P^2} 
\, + \,
  k_\alpha     S L_\alpha H     \frac{\langle W\rangle}{M_P^3}
\, + \,
  \frac{1}{2}h S S              \frac{\langle W\rangle^2} {M_P^5}\,.
\label{spottwo}
\end{equation}
$\langle W\rangle$ is here a constant term of the superpotential. It
carries $R$-charge {\it two} and has value $m_{3/2} M_P^2$ in order to
cancel the vacuum energy density arising from the supersymmetry
breaking sector~\cite{Nilles:1984ge}. Notice that the mixed mass terms
$\bar{N}_\alpha S$, as the bilinear Higgs term $H \bar{H}$, are
naturally at the scale $m_{3/2}$. They give rise to mixed fermion mass
terms $\bar{\nu}_{R\,\alpha} \nu_{L s}$ and 
$\bar{\nu}_{L s}\nu_{R\,\alpha}$ for right-handed and sterile
neutrino current eigenstates, where $\nu_{Ls}$ and $\nu_{R\,\alpha}$ 
are respectively the
fermionic component of $S$ and the charged conjugated fermionic
components of the superfields $\bar{N}_\alpha$. As usual, the
fermionic components of the neutrino superfields in the doublets
$L_\alpha$ are denoted by $\nu_{L\alpha}$. (It is understood here that
the charged lepton fields are in the basis in which their mass matrix
is diagonal.)

Thus, on the basis $\{\nu_{Ls},\nu_{L\alpha},\nu^c_{R\alpha}\}$, the 
$7 \times 7$ neutrino mass matrix gets the form:
\begin{equation}
\left(
\begin{array}{c|cc}
 h \epsilon m_{3/2} & {\sf k}^T \epsilon v  & {\sf f}^T m_{3/2} \\[1.01ex]
\hline  & &                                                     \\[-2.ex]
 {\sf k} \epsilon v & {\bf  0}              & {\bf y}^T v       \\
 {\sf f} m_{3/2}    & {\bf y}  v            & {\bf M_R}         \\
\end{array}
\right) \,,
\end{equation}
when a matrix notation is adopted: ${\bf M_R}$ and ${\bf y}$ are
$3\times 3$ matrices in generation space, ${\sf f}$ and ${\sf k}$ are
$3$-component vectors. The dimensionless parameter $\epsilon$ is the
ratio $m_{3/2}/M_P$. In the limit $\epsilon \to 0$, this matrix is of
rank {\it six} and one eigenvalue vanishes identically. The actual
neutrino spectrum can be easily read from the mass matrix for the
four light neutrinos, obtained by integrating out the three heavy
ones, of mass $\sim M_R$. This $4\times 4$ mass matrix has the form:
\begin{equation}
- \left(
\begin{array}{c|c}
 {\sf f}^T \displaystyle{\frac{1}{\bf M_R}}{\sf f} \,m_{3/2}^2 
 - h \epsilon m_{3/2}                                           &
 {\sf f}^T \displaystyle{\frac{1}{\bf M_R}}{\bf y} \,m_{3/2} v
 - {\sf k}^T \epsilon v                                         \\[1.8ex]
  \hline  &                                                     \\[-2.ex]
 {\bf y}^T \displaystyle{\frac{1}{\bf M_R}} {\sf f} \,m_{3/2} v
 - {\sf k} \epsilon v       & 
 {\bf y}^T \displaystyle{\frac{1}{\bf M_R}} {\bf y} v^2             \\
\end{array}
\right)\,.
\label{matrix44}
\end{equation}

The spectrum of light neutrinos is then composed of two
states, $\nu_2$ and $\nu_3$, with mass $\sim (y v)^2 /M_R$; one,
$\nu_s$, at the scale $\sim (f m_{3/2})^2/M_R$, and one, $\nu_1$, with 
much smaller mass $\sim v^2/M_P$.  Since $m_{3/2}$ is in general
expected to be larger than the electroweak scale 
($v\simeq {\cal O}(100)\,$GeV) by roughly one order of magnitude, the
hierarchy in the ``$1\!+\!3$'' scheme between $m_{\nu_s}$ and the
heavier of the two states $\nu_2$ and $\nu_3$, say $\nu_3$, is
naturally obtained. For $M_R \sim 10^{15}\,$GeV, it is 
$m_{\nu_s} \sim 1\,$eV and $m_{\nu_3} \sim 10^{-2}\!-\!10^{-1}\,$eV.
Without much tuning of the couplings $y$ and the masses $M_R$, the
lighter state in this pair can then be given a mass 
$m_{\nu_2}^2 \sim \Delta m^2_{sol}$. For the MSW solution to the solar
neutrino problem, $m_{\nu_2}$ needs to be only about one order of
magnitude smaller than $m_{\nu_3}$.

Furthermore, it is easy to see from the matrix~(\ref{matrix44}), that
the mixing angles between the sterile neutrino and other three active
neutrinos is of order of $v/m_{3/2}\sim 0.1$, which is the correct
order of magnitude for the ``$1\!+\!3$'' scheme. This implies,
together with $m_{\nu_s}\sim 1\,$eV, that the neutrino mass term
$m_{\nu_e\,\nu_e}$ contributing to the $2\beta$ decay is in a range of
$10^{-2}$--$10^{-3}\,$eV, which is accessible to the future $2\beta$
decay experiments~\cite{Doublebeta}.

The intermediate scale $M_R$ can be dynamically explained as the
breaking scale of an additional gauge symmetry. The simplest candidate
for this gauge symmetry is $U(1)_{B-L}$, where $B$ and $L$ are the
baryon and the lepton number, respectively.  Given the presence of the
sterile neutrino superfield $S$, another sterile neutrino superfield
$\bar{N}_o$ needs to be added, in order to cancel $U(1)_{B-L}$ gauge
anomalies. The gauge symmetry is broken by two chiral multiplets
$\Phi$ and $\bar{\Phi}$ carrying nonvanishing $U(1)_{B-L}$ charges.
The $U(1)_{B-L}$ charges for the fields $L_\alpha$, $H$, $\bar{H}$,
$\bar{N}_\alpha$ is uniquely fixed (up to a $U(1)_Y$ transformation) 
by the requirement of exactly vanishing gauge anomalies as follows:
\begin{equation}
 X(L_\alpha)         = -1, \ \, 
 X(H)                =  0, \ \, 
 X({\bar{H}})        =  0, \ \, 
 X({\bar{N}_\alpha}) = +1  \,; 
\label{bmlcharges_one}
\end{equation}
we choose those for $\bar{N}_o$, $S$, $\Phi$, and $\bar{\Phi}$ to be:
\begin{equation}
 X(\Phi)         = +2, \ \, 
 X({\bar{\Phi}}) = -2, \ \,
 X({\bar{N}_o})  = +1, \ \,
 X(S)            = -1\,.
\label{bmlcharges_two}
\end{equation}
Finally, we assign $R$-charges to the additional superfields
$\bar{N}_o$, $\Phi$, and $\bar{\Phi}$ according to the standard
prescription for fermionic matter and Higgs superfields:
\begin{equation}
  R(\bar{N}_o)  =  1, \quad 
  R(\Phi)       =  0, \quad 
  R(\bar{\Phi}) =  0\,. 
\label{rcharges_two}
\end{equation}
(Notice that the quantum numbers for the additional $\bar{N}_o$ 
are the same as those for $\bar{N}_\alpha$.) 
The two terms $W_1$ and $W_2$ in the superpotential are now modified 
as follows. $W_1$ has the form:
\begin{equation}
 W_1 = 
  y_{\alpha\beta} \bar{N}_\alpha L_\beta  H                          \, + \,
  y_{o \alpha}    \bar{N}_o  L_\alpha     H                          \, + \,
 \frac{1}{2} z_{\alpha\beta}  \bar{\Phi}\bar{N}_\alpha \bar{N}_\beta \, + \,
             z_{o  \alpha}    \bar{\Phi}\bar{N}_o      \bar{N}_\alpha\, + \,
 \frac{1}{2} z_{oo}           \bar{\Phi}\bar{N}_o      \bar{N}_o \,,
\label{spotonenew}
\end{equation}
where 
\begin{equation}
  z_{\alpha\beta} \langle v_{\bar{\Phi}} \rangle =  M_{R\,\alpha\beta}, 
\quad 
  z_{o \alpha}    \langle v_{\bar{\Phi}} \rangle =  M_{R\,o \alpha}, 
\quad 
  z_{o o}         \langle v_{\bar{\Phi}} \rangle =  M_{R\,o o}\,. 
\end{equation}
$W_2$ is modified to include an additional operator mixing 
the superfields $\bar{N}_o$ and $S$, and to have a more suppressed 
form for the last two operators:
\begin{equation}
 W_2 =  
  h_H            H \bar{H}        \frac{\langle W\rangle}{M_P^2}   
\, + \,
  f_\alpha      \bar{N}_\alpha S  \frac{\langle W\rangle}{M_P^2} 
\, + \,
  f_o           \bar{N}_o S       \frac{\langle W\rangle}{M_P^2} 
\, + \,
  k_\alpha      \Phi S L_\alpha H \frac{\langle W\rangle}{M_P^4}
\, + \,
  \frac{1}{2}h  \Phi S S          \frac{\langle W\rangle^2}{M_P^6}\,,
\label{spottwonew}
\end{equation}
as required by the $U(1)_{B-L}$ charge 
assignment~(\ref{bmlcharges_one}) and~(\ref{bmlcharges_two}).

Again, going to a matrix notation, the $8 \times 8$ neutrino mass
matrix for the eight states
$\{\nu_{Ls},\nu_{L\alpha},\nu^c_{R\alpha},\nu^c_{R o}\}$ is now:
\begin{equation}
\left(
\begin{array}{c|ccc}
 h \epsilon \epsilon^\prime m_{3/2}         & 
 {\sf k}^T \epsilon \epsilon^\prime v       & 
 {\sf f}^T m_{3/2}  & {f_o} m_{3/2}         \\[1.01ex]
\hline  & &                                 \\[-2.ex]
 {\sf k} \epsilon \epsilon^\prime v         &  
 {\bf  0}                                   & 
 {\bf y}^T v        &  {\sf y_o} v          \\
 {\sf f} m_{3/2}    &  {\bf y} v            &  
 {\bf M_R}          &  {\sf M_R}            \\
 {f_o} m_{3/2}      &  {\sf y_o}^T v          & 
 {\sf M_R}^T        &   M_R                 \\  
\end{array}
\right) \,,
\end{equation}
where, with an abuse of notation, the same symbol is used to indicate
a $3\times 3 $ matrix ${\bf M_R}$, a $3$-component vector 
${\sf M_R}$, and a $1$-component number $M_R$. The symbol 
${\sf y_o}$ indicate the $3$-component vector collecting all
couplings $y_{o\alpha}$, and $\epsilon^\prime$
is the additional suppression factor 
$\langle v_{\bar{\Phi}}\rangle/M_P$. In this case, differently from
the case without $\bar{N}_o$, the rank of the matrix remains the 
maximal one, i.e.{\it eight}, even in the limit 
$\epsilon\epsilon'\to 0$.  Once the heavy states are integrated out,
the $4\times 4$ mass matrix for the light states is formally of 
the same type as the matrix in eq.~(\ref{matrix44}). Its exact 
form is:
\begin{equation}
- \left(
\begin{array}{c|c}
  \left({\sf f}^T,\,f_o\right)
  \displaystyle{\frac{1}{\bf {\cal M}}}
  \left(\!\!\!\begin{array}{c} {\sf f}\\ f_o \end{array}\!\!\!\right)
  \,m_{3/2}^2 
  - h \epsilon\epsilon' m_{3/2}                      
&
  \left({\sf f}^T,\,f_o\right)
  \displaystyle{\frac{1}{\bf {\cal M}}}
  \left(\!\!\!\begin{array}{c} {\bf y}\\ {\sf y_o}^T \end{array}\!\!\!\right)  
  \,m_{3/2} v
  - {\sf k}^T\epsilon\epsilon' v   
\\[1.8ex]
\hline  &                                                 \\[-2.ex]
  \left({\bf y}^T,\,{\sf y_o} \right)
  \displaystyle{\frac{1}{\bf {\cal M}}}
  \left(\!\!\!\begin{array}{c}{\sf f}\\ f_o \end{array}\!\!\!\right)
  \,m_{3/2} v
  - {\sf k} \epsilon\epsilon' v       
& 
  \left({\bf y}^T,\,{\sf y_o} \right)  
  \displaystyle{\frac{1}{\bf {\cal M}}}
  \left(\!\!\!\begin{array}{c}{\bf y}\\ {\sf y_o}^T \end{array}\!\!\!\right)  
\end{array}
   \right)\,,
\end{equation}
where the matrix ${\bf {\cal M}}$ is the $4\times 4$ matrix of 
heavy masses:
\begin{equation}
  {\bf {\cal M}} = 
  \left(
   \begin{array}{cc}
    {\bf M_R} & {\sf M_R}
     \\
    {\sf M_R}^T & M_R
   \end{array}
   \right)\,.
\end{equation}
The mass spectrum and the mixing angles
are similar to those obtained in the case without $\bar{N}_o$,
although in this case $m_{\nu_1}$ is not suppressed with respect to
$m_{\nu_2}$ and $m_{\nu_3}$.

Notice that, because of the large scale $M_R$, no large radiative
contributions to neutrino masses, as those described in
Ref.~\cite{Borzumati:2000mc}, can arise. Moreover, given the
$U(1)_{B-L}$ and the $U(1)_R$ charge assignments, trilinear
contributions to the scalar potential, involving the scalar component
of the superfield $S$, $\tilde{S}$, as well as the bilinear term 
$B_S \tilde{S}\tilde{S}$ are suppressed. Therefore, also radiative
contributions mediated by $\tilde{S}$~\cite{BHNY} do not alter the
tree-level spectrum given above.

This model may be easily extended to a GUT scheme, with gauge
group $SU(5)_{\rm GUT}\times U(1)_5$. The additional $U(1)_5$ is the
so-called fiveness. By breaking $SU(5)_{\rm GUT}\times U(1)_5$ down to
the SM, one ends up with the group 
$SU(3)\times SU(2)\times U(1)_Y\times U(1)_{B-L}$, where $B-L$ is
given by a linear combination of the fiveness $Y_5$ and hypercharge
$Y$:
\begin{equation}
 B-L = \frac{1}{5}Y_5 + \frac{4}{5}Y \,.
\end{equation}
The usual quarks, leptons and neutrinos $\bar{N}_\alpha$ transform as
$({\bf 5}^*,-3)$, $({\bf 10},+1)$, and $({\bf 1},+5)$, while the
additional $\bar{N}_o$ and the sterile neutrino $S$ as $({\bf 1},+5)$
and $({\bf 1},-5)$. Here the numbers in the parentheses denote the
dimensions of $SU(5)_{\rm GUT}$ representations and the fiveness $Y_5$
charges. However, a further unification to a $SO(10)$ or $E_6$ model
seems difficult, due to the presence of the two additional fields
$\bar{N}_o$ and $S$.  One may consider that the presence of these
additional fields is explained by more fundamental theories (e.g.,
superstring theories) beyond GUTs.

\vspace*{1truecm}
\noindent 
{\bf Acknowledgements}\\ 
F.B. is supported by a Japanese Monbusho fellowship and thanks the KEK
theory group, in particular Y.~Okada, and the Tokyo University theory
group for hospitality.  The work of K.H. was supported by the Japanese
Society for the Promotion of Science. T.Y. acknowledges partial
support from the Grant-in-Aid for Scientific Research from the
Ministry of Education, Sports, and Culture of Japan, on Priority Aerea
\# 707: ``Supersymmetry and Unified Theory of Elementary Particles''.

\end{document}